\documentclass[]{article}
\usepackage{authblk}
\usepackage{xcolor}
\usepackage{amsmath}
\usepackage{amsfonts}
\usepackage{mathtools}
\usepackage{xfrac}
\usepackage{wrapfig}
\usepackage{caption}
\usepackage[margin=1in]{geometry}

\usepackage{enumitem}
\usepackage{multirow}

\renewcommand{\i}{{\mathrm{i}}}
\renewcommand{\d}{{\mathrm{d}}}
\newcommand{\e}{{\mathrm{e}}}

\begin{document}

\title{Can Efficient Fourier-Transform Techniques Favorably Impact on Broadband Computational Electromagnetism?}

\author{Thomas G. Anderson\thanks{Department of Computational Applied Mathematics and Operations Research, Rice University, Houston 77098 USA\\
$^\dagger$Department of Mathematics \& Statistics, University of New Hampshire, Durham 03824 USA\\
$^{\S}$LSEC, Institute of Computational Mathematics and Scientific/Engineering Computing, Academy of Mathematics and Systems Science, Chinese Academy of Sciences, Beijing 100190, China\\
$^{\sharp}$Department of Computing and Mathematical Sciences, California Institute of Technology, Pasadena 91125 USA}, Mark Lyon$^\dagger$, Tao Yin$^{\S}$ and Oscar P. Bruno$^{\sharp}$}

\maketitle

\vspace{-.3in}
\begin{abstract}
    In view of recently demonstrated joint use of novel Fourier-transform techniques and effective high-accuracy frequency domain solvers related to the Method of Moments, it is argued that a set of transformative innovations could be developed for the effective, accurate and efficient simulation of problems of wave propagation and scattering of broadband, time-dependent wavefields. This contribution aims to convey the character of these methods and to highlight their applicability in computational modeling of electromagnetic configurations across various fields of science and engineering.

\smallskip
\noindent \textbf{Keywords.} Time-frequency hybrid methods, time-domain scattering, Fourier transform, integral equations, dispersive media.
\end{abstract}

\maketitle

\section{Introduction}
The simulation of broadband time-transient wave propagation permeates the field of electrical engineering, and, indeed, much of engineering and science as a whole, as it impacts on applications as varied as radar, TV and radio broadcasting, wireless connectivity, signal processing, ultrasound medical imaging,  electromagnetic compatibility, and terrestrial and satellite communications (including  cellular and fiber optics networks), among many others. In all cases these technologies rely on use of an electromagnetic (or acoustic) ``signal''  that changes continuously in a time-dependent fashion: the signal's amplitude, frequency, and phase can (and typically do) all vary with time. The broadband character of signals enables AM and FM modulation, it facilitates simultaneous transmission of large amounts of data, it endows high resolution to radar systems for remote sensing and weather monitoring, and it underlies a myriad of everyday applications of great importance. This contribution suggests that the use of a novel set of Fourier-transform techniques in conjunction with effective frequency domain solvers related to the Method of Moments can lead to significant advances in the computational solution of broadband propagation and scattering problems underlying many or all of these areas.

The Finite-Difference Time-Domain method~\cite{Taflove:00} (FDTD) is the recognized work-horse for the simulation of time-dependent wave propagation and scattering: it provides a simple, broadly applicable and highly effective approach for the simulation of time-transient signals. Finite Element Time Domain (FETD) methods~\cite{Lee:97}, in turn, offer substantial geometric flexibility, though at the cost of increased computational demands and implementation complexity. These approaches, which proceed via discretization of the spatio-temporal propagation domain (e.g., via finite-difference approximations on equispaced grids for FDTD), and which additionally require the use of absorbing boundary conditions to simulate unbounded-space propagation~\cite{Bayliss:80,Berenger:94}, do present certain challenges---most notably, finite difference- and finite element-based methods suffer from numerical dispersion, necessitating the use of fine spatial meshes for accuracy, and consequently fine temporal meshes to maintain stability---ultimately limiting their performance for high-frequency or long-range simulations.

Green function-based methods (GF), in turn, are best known for their efficiency and accuracy, at least for frequency domain applications (GFFD); as discussed in what follows, a number of corresponding time-domain versions (GFTD) have also been proposed~\cite{HaDuong:86,HaDuong:03,Yilmaz:04,Barnett:20}, albeit with mixed success on account of the challenging light-cone integration that they require and the difficulties  that have been observed in regard to their temporal stability. Green function-based algorithms operate by expressing the electromagnetic field as a ``linear combination'' (or more properly, as a result of integration) of certain exact solutions that are given, precisely, by the Green function. Thus, Green function methods achieve as close an approximation, as has thus far proven possible, to the ideal of obtaining closed-form solutions for general structures. The enforcement of boundary conditions in the frequency-domain case does necessitate the solution of a system of equations for what are typically referred to as the ``surface currents''. Although the dimensionality of the Green-function linear system is reduced by one when compared to their volumetric counterparts (since unknowns are only placed on scattering surfaces and not on the propagation volume), the matrices arising from methods based on Green functions are dense (while those associated with the finite-difference and finite-element counterparts are sparse), which presents a challenge. But, fortunately, a variety of acceleration methods introduced over the last few decades have greatly mitigated the challenge~\cite{cheng2006wideband, michielssen1994multilevel, bleszynski1996aim, phillips1997precorrected, bruno2001fast, engquist2007fast, bauinger2021interpolated} to the point that the efficiency of existing GFFD methods for fixed frequency problems is broadly understood and accepted.

As mentioned above, a variety of GFTD methods~\cite{HaDuong:86,HaDuong:03,Barnett:20} have been proposed over the last few decades in which time-domain Green functions are used as a ``basis'' of solutions, from which all time-domain solutions can be obtained by integration. As in the frequency domain case, accelerated GFTD algorithms have been proposed~\cite{Yilmaz:04} to mitigate the cost of the evaluation of pairwise interactions. As is well known, however, GFTD formulations incorporate the Dirac delta function that is part of the time-domain Green function, and thus give rise to integration domains equal to the intersection of the light cone with the overall scattering surface, and they do require time-stepping. Thus, GFTD approaches generally result in complex schemes for which, additionally, ensuring stability has been rather challenging, and they have typically been implemented in low-order accuracy setups and, thus, with significant numerical  time-stepping dispersion error. High-order versions have recently been proposed but have proven particularly expensive in terms of computing time and memory (see, in particular, Table~\ref{table} for comparisons of the proposed method to previous contributions).

In the context of this article it is interesting to recall that the time-domain Green function coincides with the inverse Fourier transform, from frequency to time, of the frequency-domain Green function: the closed-form time-domain Green function expression mentioned above, which incorporates a Dirac delta function, results from exact evaluation of the corresponding Fourier transform integral. Clearly, then, alternative approaches could be pursued in which time-domain solutions are sought as Fourier transforms of frequency-domain solutions---e.g., by computing Fourier transforms of solutions obtained by frequency-domain Green function methods (which ultimately result in what we call Frequency-Time Hybrids in what follows) instead of directly using the closed form time-domain Green function. While this route may seem suboptimal (as the {\em closed-form} time domain Green function is discarded to rely instead on {\em numerical} Fourier transformation of frequency-domain solutions), the approach does not suffer from the aforementioned GFTD challenges. However, only a few such Fourier-transform efforts had previously been attempted, in most cases involving simple scatterers, low frequencies and short pulses---on account of the challenges posed by necessary evaluation of Fourier transforms for long times and associated high-frequency integrals with respect to frequency. Such integrals require, in particular, fine integration grids in the frequency variable that need to be made finer and finer as the desired simulation time grows, thus requiring larger and larger numbers of frequency-domain solutions---which had previously forestalled the viability of Fourier-transform methods.

Fourier/Laplace-transform methods based on discrete analogs of the Laplace transformation that proceed in conjunction with time-stepping schemes, which are known as convolution quadrature methods (CQ), have also been proposed. Like all Fourier-transform methods the CQ methods benefit from their reliance on decoupled frequency domain problems which can be solved in an embarrassingly parallel fashion. Unfortunately, however, in view of the time-stepping stage that they utilize, these approaches incur temporal dispersion errors~\cite{Banjai:12}, while the Laplace inversion stage additionally raises delicate stability questions~\cite{betcke2017overresolving}. Most notably, further, the discrete Laplace transformation destroys the finite-time history dependence enjoyed by the FDTD methods by introducing a certain ``infinite tail'' of dependence~\cite{Sayas:16}, which makes the incremental time evolution increasingly expensive as time grows---so that, in all, as for GFTD methods and other Fourier-transform methods, only demonstrations for brief pulses have been reported.

One of the main motivations of the present contribution lies in the recent work~\cite{Anderson:20,BrunoYin24}, which shows that these difficulties may be circumvented by means of Fourier-transform approaches that incorporate three main elements, namely, {\em windowing, recentering and high-frequency integration}. In the resulting Fast-Hybrid methods~\cite{Anderson:20,BrunoYin24} (FHM), the windowing portion of the algorithm smoothly partitions the excitation function to produce an equivalent sequence of time-domain problems whose frequency content can be evaluated, on the basis of the aforementioned recentering step, by relying on the use of a fixed (time-independent) set of frequencies; the high-frequency integration algorithm, which accurately produces integrals on the basis of fixed integration meshes for arbitrarily high frequencies, then completes the basic elements of the new FHM.

As illustrated in Section~\ref{sec:fhm_outline}, the resulting FHM algorithms can be quite effective, and they are endowed with a number of interesting properties: they can tackle complex physical structures, they enable parallelization in time in a straightforward manner, they can natively handle dispersive media, and they allow for time leaping---that is, solution sampling at any given time $T$ at $O(1)$-bounded sampling cost, for arbitrarily large values of $T$, and without requirement of evaluation of the solution at intermediate times. As illustrated by a number of numerical examples presented in this article, further, the method is well suited to tackle complex geometries, and it can perform significantly better than most alternative approaches. It is expected that, combined with state-of-the-art frequency-domain solvers the novel Fourier-transform strategy will provide a highly competitive, accurate and efficient approach for the solution of a scientific and engineering centerpiece: the broadband propagation and scattering problem.

\section{Efficient frequency-time hybrids via time partitions of unity}\label{sec:fhm_outline}
We consider the exterior scattering boundary value problem for the time-dependent wave equation
\begin{subequations}\label{w_eq}
    \begin{align}
        \frac{\partial^2 u}{\partial t^2}(\mathbf{r}, t)& - c^2\Delta u
            (\mathbf{r}, t) = 0,\quad\mathbf{r} \in \Omega,\label{w_eq_a}\\
        u(\mathbf{r},0) &= \frac{\partial u}{\partial t}(\mathbf{r}, 0)
            = 0,\quad\mathbf{r} \in \Omega,\\
        u(\mathbf{r}, t) &=  b(\mathbf{r}, t)
            \quad\mbox{for}\quad(\mathbf{r},t)\in\Gamma\times [0,T^\textit{inc}],\label{w_eq_c}
    \end{align}
\end{subequations}
and Maxwell system
\begin{subequations}\label{maxwell}
\begin{align}
    \nabla \times \mathbf{E} &= -\frac{\partial \mathbf{B}}{\partial t}\\
    \nabla \times \mathbf{B} &= \mu_0 \mathbf{J} + \mu_0 \epsilon_0 \frac{\partial \mathbf{E}}{\partial t}\\
    n \times \mathbf{E} &= - n \times \mathbf{E}^\textit{inc} \quad\mbox{ on } \quad \Gamma.\label{maxwell_c}
\end{align}
\end{subequations}
In~\eqref{w_eq_c}, $b$ is a given function, such as e.g.\ a plane-wave which impinges on $\Gamma$, while in~\eqref{maxwell_c} $\mathbf{E}^\textit{inc}$ is an incident electric field. For the sake of simplicity, in what follows the FHM is described for the case of the acoustic equations~\eqref{w_eq}; numerical illustrations for the Maxwell equations are also presented in this contribution. For the FHM implementation in the acoustic case it is necessary to consider the frequency-domain counterpart of~\eqref{w_eq}, namely, the Helmholtz equation
\begin{subequations}\label{helmholtz}
    \begin{align}
        &\Delta U(\mathbf{r}, \omega) + \kappa^2(\omega)U(\mathbf{r}, \omega) = 0, \quad \mathbf{r} \in
        \Omega,\\
        &U(\mathbf{r}, \omega) = B(\mathbf{r}, \omega), \quad \mathbf{r} \in \Gamma,
    \end{align}
\end{subequations}
where $U$ denotes the Fourier transform, with respect to time of the solution $u$, and where $\kappa(\omega) = \omega / c$ (but see also Section~\ref{sec:dispersive} where dispersion relations corresponding to dispersive media are considered).

\begin{figure}
    \centering
    \includegraphics[width=0.49\textwidth]{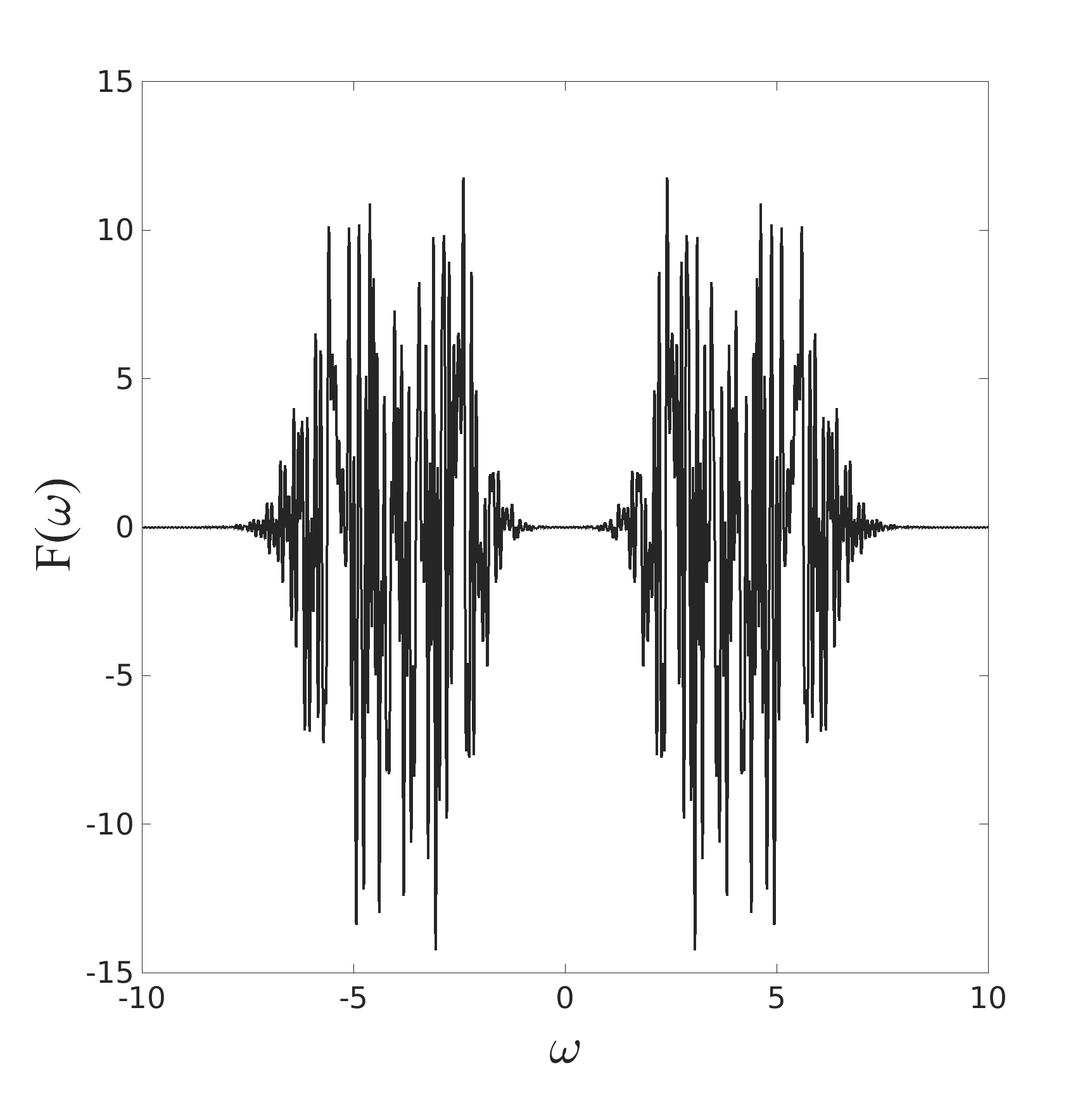}
    \includegraphics[width=0.49\textwidth]{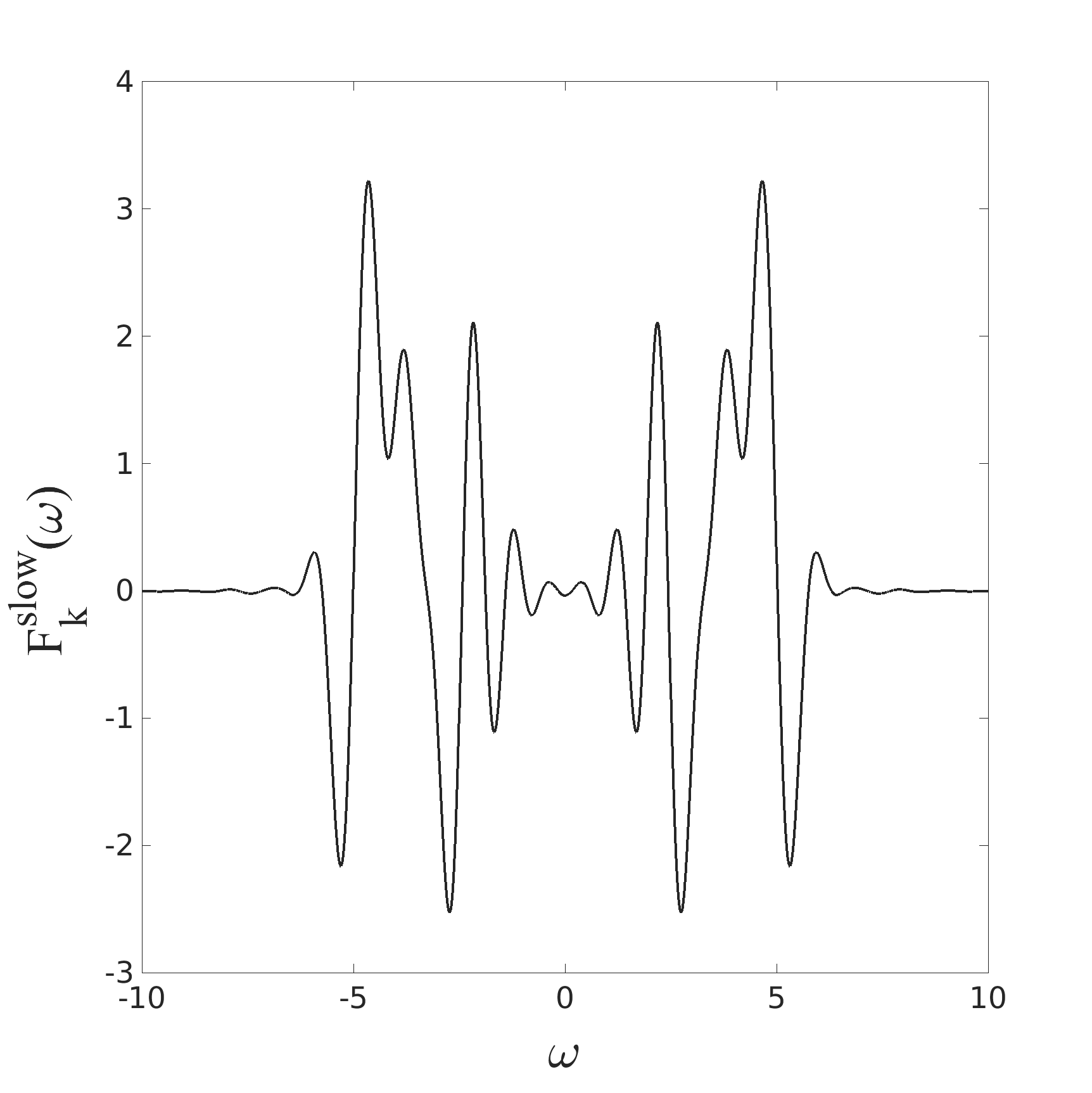}
    \caption{Left: real part of the Fourier transform $F(\omega)$ of $f(t)$. Right: real part of the
       windowed and recentered Fourier transform $F_k^\textit{slow}(\omega)$ given in~\eqref{Fkslow_int}. Unlike the Fourier transform of $f(t)$, the Fourier transform of the windowed and recentered function $f_k(t) = w(t+s_k)f(t+s_k)$ is slowly oscillatory with respect to $\omega$.}
    \label{fig:Ms1}
\end{figure}

Throughout this paper the nomenclature
\begin{equation}\label{transform_pair}
    F(\omega) =\int_{-\infty}^{\infty} f(t) \e^{\i
    \omega t}\, \d t,\qquad
    f(t) = \frac{1}{2\pi} \int_{-\infty}^\infty F(\omega) \e^{-\i \omega t}\, \d \omega
\end{equation}
is used to denote Fourier-transform pairs such as $(f,F)$. For example, in the context of equations~\eqref{w_eq} and~\eqref{helmholtz}, for each fixed $\mathbf{r}$ we have the Fourier transform pairs
\begin{equation}
    B(\mathbf{r}, \omega)\leftrightarrow b(\mathbf{r}, t) \qquad\mbox{as well as}\qquad U(\mathbf{r}, \omega)\leftrightarrow u(\mathbf{r}, t).
\end{equation}
While, as mentioned above, any discretization method for the continuous Fourier transform pair would yield a solver for~\eqref{w_eq}, such a  direct approach does not result in an efficient solver. To understand this, consider a smooth signal \( f(t) \) that vanishes outside the time interval \([0, T]\) for sufficiently large $T$, and assume $f(t)$ does not vanish for large values of $t$; the Fourier transform $F(\omega)$ is shown in the left panel of Figure~\ref{fig:Ms1}. Since $f(t)$ does not vanish for large values of $t$, the integrand in the left integral in~\eqref{transform_pair} oscillates rapidly as a function of $\omega$, at least for the existing large $t$ values, and therefore so does the integral $F(\omega)$. In the context of the scattering problem~\eqref{w_eq} we have $F(\omega) = U(\mathbf{r}, \omega)$, and the fast oscillations observed would require the use of fine discretization meshes in the $\omega$ frequency domain in order to produce $u(\mathbf{r}, t)$, which proves an extremely costly procedure, since for each such frequency a solve of the Helmholtz equation~\eqref{helmholtz} is needed. This is one of the two historically limiting factors that have hindered the use of Fourier transform strategies for the solution of scattering problems.

The second such difficulty in the computation of the time-domain solution $u(\mathbf{r}, t)$ concerns the required evaluation of an inverse Fourier transform akin to the second integral in~\eqref{transform_pair}, which as just mentioned, contains a highly-oscillatory integrand.
Previous Fourier-transform methods discretize the Fourier transform integral using the trapezoidal quadrature rule:
\begin{equation}
    u(\mathbf{r}, t) \approx
    \frac{1}{2\pi} \int_{-W}^{W}
    U(\mathbf{r}, \omega) \e^{-\i t \omega}\,\d\omega \approx \frac{W}{2\pi m}\sum_{k=0}^{m-1} U(\mathbf{r}, \omega_k)
    \e^{-\i t\omega_k} \quad \left(\omega_k = -W + k\, \Delta \omega\right).
\label{trap_rule_four_trans}
\end{equation}
As is well known, the evaluation of this discrete Fourier transform  can be performed for many values of $t$ by means of the FFT algorithm, at a $\mathcal{O}(N\log N)$ computing cost, where $N$ denotes the number of $t$ values at which the solution is to be computed. But, unfortunately, in the present context more relevant than this advantage is its significant flaw: for large $t$ values, spurious periodicity renders the approximation in~\eqref{trap_rule_four_trans} invalid. The onset of spurious periodicity can be delayed by utilizing finer frequency grids (i.e.\ with smaller $\Delta \omega$ values), but as before this requires the computation of a number of additional Helmholtz solutions that \emph{grows linearly with evaluation time $t$}.

The proposed group of three ideas (windowing, re-centering, and high-frequency integration) tackles both of these challenges, ultimately yielding a method that relies only on coarse frequency grids. The first idea deals with the problem posed by a long time-duration incident field (whose Fourier transform is a signal akin to that displayed in the left panel of Figure~\ref{fig:Ms1}) by splitting the boundary data $b(\mathbf{r}, t)$, assumed to vanish outside the interval $[0, T]$ for $\mathbf{r} \in \Gamma$,  as a sum of sequence of windows $b_k(\mathbf{r}, t)$, each one of which is nonzero for a small time span. In detail, to achieve such partitioning, we lay down a time partition of unity (POU) set $\mathcal{P} =
\{w_k(t)\,|\,k=1,\dots,K\}$ of windowing functions, where $w_k(t)$ vanishes outside a neighborhood of the time $t=s_k$ for
a window center $s_k\in [0,T]$ ($1\leq k\leq K$). For an adequately selected value $H>0$ the partition of unity window functions satisfy certain conditions, namely
\begin{enumerate}[label=\alph*)] 
    \item $s_{k+1} - s_k =  3H/2$,
    \item $w_k(t)=1$ in a neighborhood $|t - s_k| < H/2$,
    \item $w_k(t)=0$ for $|t- s_k|>H$, and
    \item\label{three} $\sum_{k=1}^K w_k(t) = 1\quad \mbox{for all}\quad t\in
        [0,T]$.
\end{enumerate}
Note that, since $H$ is
$T$-independent, the integer $K$ is necessarily an $\mathcal{O}(T)$
quantity. Several approaches may be used to leverage the partition of unity to produce windowed data \( b_k \) required by the FHM. The simplest such approach is to set \( b_k(\mathbf{r},t) = w_k(t) b(\mathbf{r},t) \). However, following the method of Anderson~\cite{Anderson:20}, we instead apply the partition of unity in a similar manner but to a scalar ($\mathbf{r}$-independent) signal function $a=a(t)$ which is used to define the frequency dependent amplitude of the incident field $b(\mathbf{r}, t)$. In any case, the windowed data satisfies $b(\mathbf{r}, t) = \sum_{k=1}^K b_k(\mathbf{r}, t)$. Each windowed data function $b_k$ leads to a solution $u_k$ satisfying $u_k(\mathbf{r}, t) = b_k(\mathbf{r}, t)$ for $\mathbf{r} \in \Gamma$.

The POU has favorable implications when used together with Fourier transforms since taking a smooth function $f(t)$ and defining $f_k(t) = w_k(t) f(t)$ we obtain
\begin{equation}\label{Fk_sum}
  F(\omega) = \sum_{k=1}^K F_k(\omega),\quad\mbox{where}\quad
  F_k(\omega)=\int_{s_k-H}^{s_k+H} f_k(t) \e^{\i \omega t}\,\d t.
\end{equation}
The re-centering component of the strategy now enters by writing the integral for $F_k(\omega)$ using the identity
\begin{equation}\label{fk_def}
F_k(\omega)= \int_{-H}^{H} f_k(t+s_k) \e^{\i \omega (t+s_k)}\,\d t=  \e^{\i \omega s_k} F_k^\textit{slow}(\omega)
\end{equation}
with
\begin{equation}\label{Fkslow_int}
  F_k^\textit{slow}(\omega)= \int_{-H}^{H} f_k(t+s_k) \e^{\i \omega t}\,\d t.
\end{equation}
The superscript ``slow'' is evocative of the fact that, in view of the limited integration range $[-H,H]$ used, the integral on the right-hand side of~\eqref{Fkslow_int} is a slowly oscillatory function of $\omega$. An example is presented in the right panel of Figure~\ref{fig:Ms1} which displays $F_k^\textit{slow}(\omega)$ corresponding to $f_k(t)$ for a particular $k$ value; the complete Fourier transform of $f$, in turn, which is displayed on the left panel of that figure, is clearly much more rapidly oscillatory.

The significance of time-windowed-and-recentered functions in the context of Fourier-transform methods can be appreciated, for example, in the ubiquitous case of a wave field $b$ propagating in a single direction $\mathbf{p}$:
\begin{equation}\label{single_incid_field}
    b(\mathbf{r}, t) = \frac{1}{2\pi} \int_{-\infty}^\infty B(\omega)
    \e^{\i \frac{\omega}{c} (\mathbf{p}\cdot\mathbf{r} - ct)}\,\d\omega\quad\mbox{with}\quad B(\omega) = \int_{-\infty}^\infty a(t) \e^{\i \omega t}\,\d t,
\end{equation}
where $a(t)$ is some smooth function which vanishes outside a bounded time interval.  Similar to~\eqref{Fk_sum} and~\eqref{fk_def} and writing $b_k(t) = w_k(t) a(t)$ and $B_k$ as the Fourier transform of $b_k$, we have
\begin{equation}\label{Bk_def}
  B(\omega) = \sum_{k=1}^K B_k(\omega),
    \quad\mbox{and}\quad B_k(\omega) = \e^{\i\omega s_k}
    B_k^\textit{slow}(\omega).
\end{equation}
In view of the linearity of the problem~\eqref{helmholtz} we may express the individual window solutions $U_k$ to the problem~\eqref{helmholtz} with $B(\mathbf{r}, \omega)$ substituted by $B_k(\omega)\e^{\i \kappa(\omega) \mathbf{p}\cdot\mathbf{r}}$ in terms of the Helmholtz solution $U_{\mathbf{p}}(\mathbf{r},
    \omega)$ with boundary values $\e^{\i \kappa(\omega) \mathbf{p}\cdot\mathbf{r}}$ which, significantly, does not involve window-specific data $B_k$, and which can therefore be re-used to construct all of the solutions $u_k$:
\begin{equation}\label{uk_rep}
  u_k(\mathbf{r}, t) = \frac{1}{2\pi} \int_{-\infty}^\infty B_k(\omega)U_{\mathbf{p}}(\mathbf{r},
    \omega)\e^{-\i\omega
    t}\,\d\omega.
\end{equation}
Further, calling
$U_k^\textit{slow}(\mathbf{r}, \omega)$ the solution
of problem~\eqref{helmholtz} with $B(\mathbf{r}, \omega)$ substituted by
$B_k^\textit{slow}(\omega)\e^{\i \kappa(\omega) \mathbf{p}\cdot\mathbf{r}}$, we obtain the representation
\begin{equation}\label{uk_rep_1}
  u_k(\mathbf{r}, t) =  \frac{1}{2\pi} \int_{-\infty}^\infty B_k^\textit{slow}(\omega) U_{\mathbf{p}}(\mathbf{r}, \omega)\e^{-\i\omega(t - s_k)}\,\d\omega,
\end{equation}
which, importantly, expresses $u_k$ in terms of a Fourier transform of a slowly-varying function of frequency $B_k^\textit{slow}$. As suggested above, this resolves one of the previously-noted major challenges in Fourier transform methods, namely, the unnecessarily-fast oscillations in certain Fourier transforms that arise merely from consideration of large times. Finally, in view of~\eqref{single_incid_field} and~\eqref{Bk_def}, the overall solution $u$ to~\eqref{w_eq} is given by
\begin{equation}\label{eq:uk_sum}
    u(\mathbf{r}, t) = \sum_{k=1}^K u_k(\mathbf{r}, t).
\end{equation}

The resolution of the second major problem mentioned above, namely, the evaluation of $u_k(\mathbf{r}, t)$ on the basis of highly-oscillatory inverse Fourier transform integrals, is provided in Section~\ref{sec:fourier_efficient}. In view of the fast decay of $U_k^\textit{slow}(\mathbf{r}, \omega)$ as $|\omega| \to \infty$ (see~\cite[Rem.\ 4]{Anderson:20}), the integral~\eqref{uk_rep_1} can be accurately approximated by
\begin{equation}\label{ukW}
    u_k(\mathbf{r}, t) \approx \frac{1}{2\pi}
    \int_{-W}^W U_k^\textit{slow}(\mathbf{r}, \omega)
    \e^{-\i\omega(t - s_k)}\,\d\omega,
\end{equation}
with errors that decrease rapidly with the truncation width $W$. Clearly the evaluation of~\eqref{ukW} requires computation of the Helmholtz solution $U_k^\textit{slow}$ at a sufficient number of frequencies, as needed for sufficiently accurate integration. The Fast-Hybrid method is agnostic to the means by which these solutions are obtained but the method has most often been applied in conjunction with frequency-domain Green function-based boundary integral formulations related to the Method of Moments~\cite{harrington1993field} for the problem~\eqref{helmholtz} (but see also Section~\ref{sec:dispersive}).

In order to illustrate the connection between the FHM and GFFD methods we consider the combined field integral equation formulation
\begin{equation}\label{CFIE_direct}
    \frac{1}{2}\Psi(\mathbf{r}, \omega) + (K^*_\omega \Psi)(\mathbf{r}, \omega) -
    \i\eta(S_\omega \Psi)(\mathbf{r}, \omega)= \frac{\partial B(\mathbf{r}, \omega)}{\partial
    n(\mathbf{r})} - \i\eta B(\mathbf{r}, \omega), \quad \mathbf{r} \in \Gamma,
\end{equation}
where $S_\omega$ and $K_\omega^*$ denote the well-known single- and adjoint-double layer operators (see~e.g.~\cite{colton1992inverse}). Once the solution $\Psi$ to this equation is obtained, the frequency-domain solution $U$ may be produced by means of a representation formula that utilizes $\Psi$ in conjunction with the Green function $G_\omega(\mathbf{r}, \mathbf{r}')$. In fact, calling $\Psi_{\mathbf{p}} = \Psi_{\mathbf{p}}(\mathbf{r}, \omega)$ the solution with data $B(\mathbf{r}, \omega)$ replaced by $\e^{\i\kappa(\omega) \mathbf{p}\cdot \mathbf{r}}$, the desired frequency-domain volumetric field $U_k^\textit{slow}$ is given by the representation formula
\begin{equation}\label{Ukslow_repr} 
\begin{split}
  U_k^\textit{slow}(\mathbf{r}, \omega) = \int_\Gamma 
  \Psi_k^\textit{slow}(\mathbf{r}', \omega) 
  & G_{\omega}(\mathbf{r}, \mathbf{r}')\,\d\sigma(\mathbf{r}') \\
  & = \underbrace{B_k^\textit{slow}(\omega)}_{\text{transient incidence}} \underbrace{\int_\Gamma
  \Psi_\mathbf{p}(\mathbf{r}', \omega) G_{\omega}(\mathbf{r},
  \mathbf{r}')\,\d\sigma(\mathbf{r}')}_{\text{window independent}},
\end{split}
\end{equation} 
where we have used the relation $U_k(\mathbf{r}, \omega) = \e^{\i\omega s_k} U_k^\textit{slow}(\mathbf{r}, \omega)$. The upshot of the windowing-and-recentering procedure is that a fixed, window-independent set of frequency domain boundary-integral solutions suffices to produce the necessary frequency-domain solutions for all windows ($k = 1, \ldots, K$) arising from the temporal POU, and it therefore only remains  to recover the time-domain solutions~\eqref{ukW}  from~\eqref{Ukslow_repr}. Clearly, to do this it is of paramount importance to utilize a numerical Fourier transformation methodology that remains efficient for arbitrarily large values of $t$; one such approach is presented in Section~\ref{sec:fourier_efficient}.

\subsection{Efficient Fourier transform evaluation}\label{sec:fourier_efficient}
To realize the full potential of the proposed Fourier-transform methods it is necessary provide a method for the evaluation of the inverse Fourier transform in~\eqref{ukW} while 1)~Avoiding the pitfalls embodied in~\eqref{trap_rule_four_trans} and mentioned immediately after that equation; and, 2)~Enabling large time evaluation at fixed cost. More generally, the evaluation of sequences of truncated direct and inverse Fourier transforms of the form 
\begin{equation}
    F(\omega) = \int_{-H}^H f(t) \e^{\i\omega t}\,\d t\quad\mbox{and}\quad f(t) = \frac{1}{2\pi} \int_{-W}^{W}
    F(\omega) \e^{-\i \omega t}\,\d\omega
\label{four_trans_limited} 
\end{equation}
is required as part of the proposed FHM algorithm, the former as part of the windowing and recentering strategy itself in~\eqref{Fkslow_int} for large $\omega$, and the second for the evaluation of~\eqref{ukW} for large $t$.
In what follows we describe the methods used by the FHM to effectively evaluate such integrals; for definiteness we do this in the case of the second integral in~\eqref{four_trans_limited}, but, of course, the first integral in that equation can be treated analogously. (Some modifications in the spatially two-dimensional case, which we omit here but which are reported in~\cite{Anderson:20}, are necessary due to possible singular behavior of e.g.\ $U_k^\textit{slow}(\mathbf{r}, \omega)$ as $\omega \to 0$.)

In order to evaluate $f(t)$ in~\eqref{four_trans_limited} the proposed method uses the Fourier series expansion
\begin{equation}\label{fourier_approx}
  F(\omega) \approx \sum_{m=-\sfrac{M}{2}}^{\sfrac{M}{2}-1} c_m \e^{\i\frac{\pi m}{W}
    \omega}
\end{equation}
which, of course, can evaluated rapidly via the FFT algorithm, 
with errors that, in view of the periodicity of the function $F(\omega)$, are super-algebraically small---i.e., smaller than any negative power of $M$. (Note that, while, strictly speaking, $F(\omega)$ is not a periodic function, in view of the near vanishing of this function and its derivatives at the limits $-W$ and $W$ of the approximate bandwidth interval, the periodic extension of $F(\omega)$ with period $2W$ can be considered, with negligible errors, as a smooth and periodic function. This function can therefore be approximated for $-W\leq \omega\leq W$ with super-algebraically small errors by its $2W$-periodic Fourier expansions up to the aforementioned near vanishing error.)  

The benefit provided by the Fourier expansion~\eqref{fourier_approx} is revealed upon its substitution into the Fourier integral, which yields
\begin{equation}\label{fourier_exp_integral}
    f(t) \approx \frac{1}{2\pi} \sum_{m=-\sfrac{M}{2}}^{\sfrac{M}{2}-1} c_m \int_{-W}^W \e^{-\i \frac{\pi}{W}(\alpha t - m)\omega}\d\omega, \quad\mbox{where}\quad \alpha = \frac{W}{\pi}.
\end{equation}
Indeed, the right-hand side of~\eqref{fourier_exp_integral} can clearly be evaluated in closed form, thus producing $f(t)$ \emph{for arbitrarily large values of $t$} with uniformly small errors that arise solely from the Fourier expansion approximation. Moreover, if applied for evaluation on a user-prescribed evaluation grid $\{t_n = n\Delta t\}_{n=N_1}^{N_2}$, the approximation in~\eqref{fourier_exp_integral} becomes the scaled discrete convolution
\begin{equation}\label{ft_scale_conv}
    f(t_n) \approx \frac{1}{2\pi} \sum_{m=-\sfrac{M}{2}}^{\sfrac{M}{2}-1} c_m b_{\beta n - m}, \quad\mbox{where}\quad b_q := 2W \operatorname{sinc}(q), \quad \beta = \alpha \Delta t.
\end{equation}
Although scaled convolutions cannot be directly accelerated with classical FFT-based algorithms for discrete convolutions, the near-convolutional structure of~\eqref{ft_scale_conv} can be exploited using the discrete fractional Fourier transform~\cite{Nascov:09,BaileySwarztrauberSIAMRev:91} to evaluate all of the desired $f(t_n)$ values in $\mathcal{O}(L \log L)$ time, where $L = \max(N_2 - N_1, M)$. In some cases arising in practice such techniques have yielded orders-of-magnitude speedups over their direct-evaluation counterparts.

\subsection{FHM illustrations for acoustic and electromagnetic problems}
\begin{figure}
    \centering
    \includegraphics[width=0.4\textwidth]{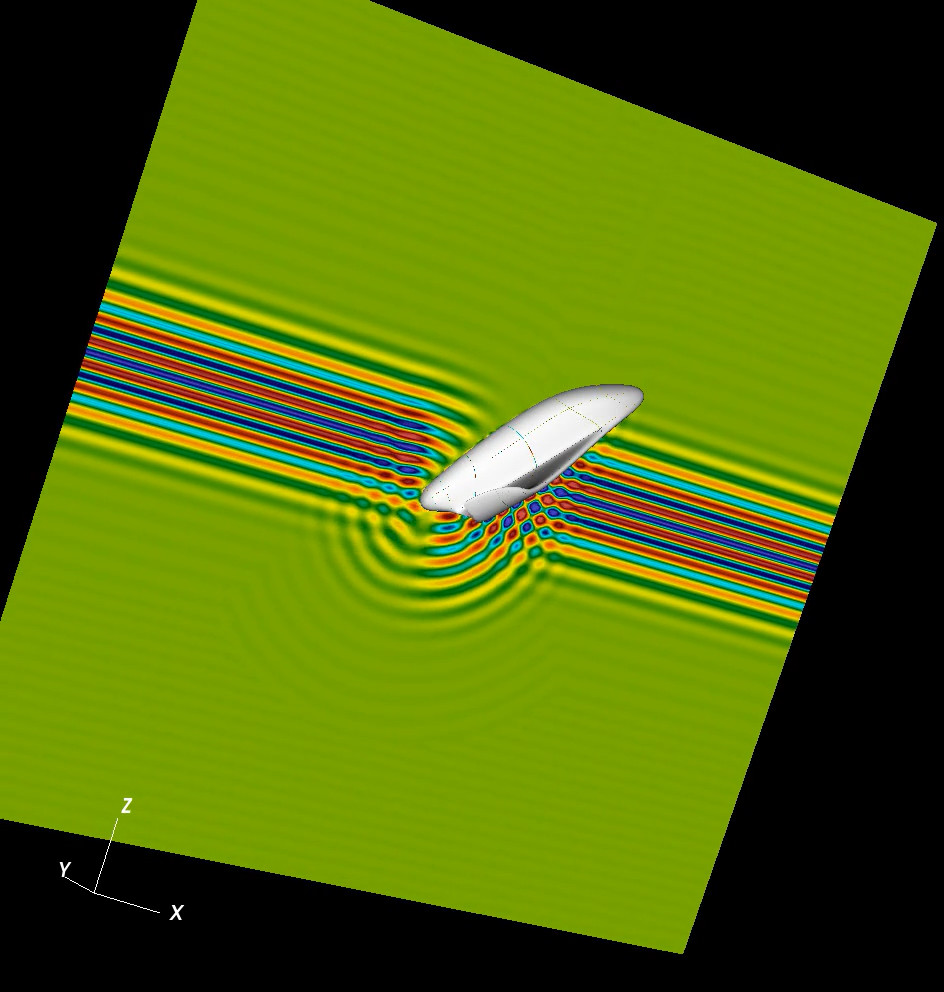}
    \includegraphics[width=0.4\textwidth]{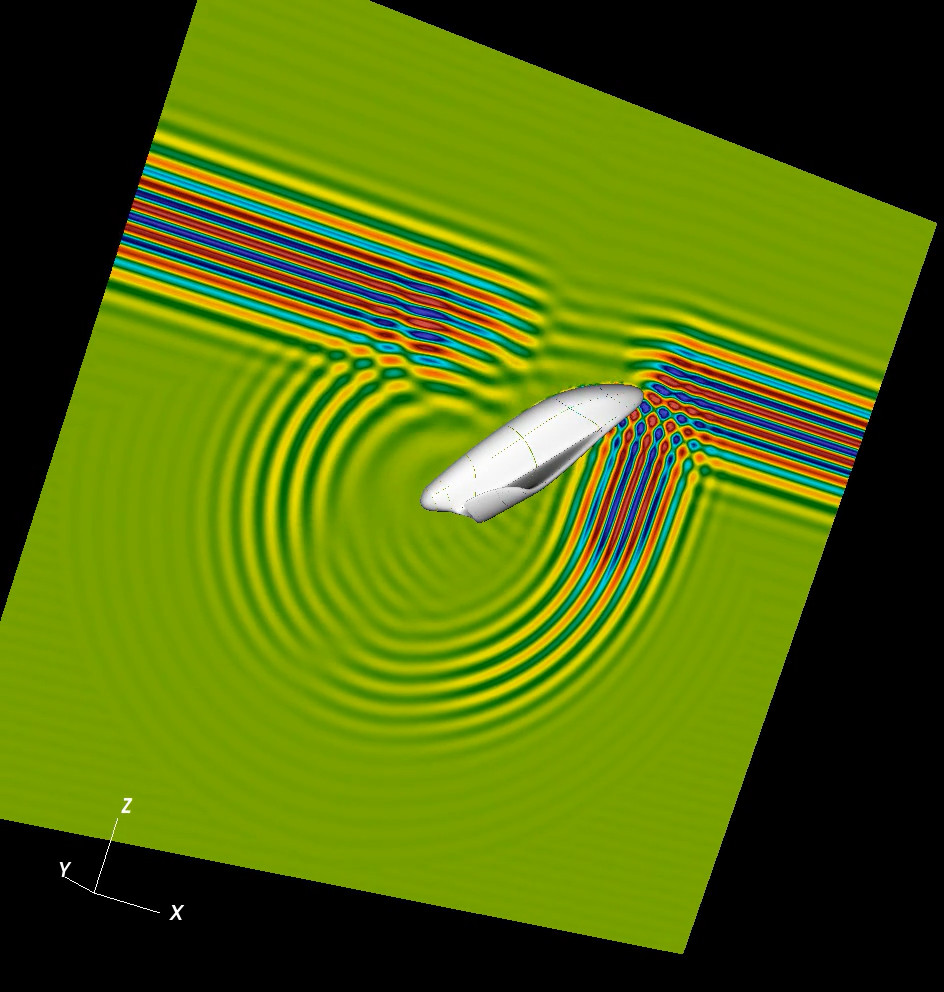}
    \caption{Scattering of an incident acoustic field with center-frequency $\omega = 15$ off of a simple aircraft structure using the FHM. A total of $250$ frequency domain integral equation solutions were used in this simulation.}
    \label{fig:glider}
\end{figure}

The overall algorithms just outlined have proven effective in the solution of a variety of acoustic and electromagnetic time-domain scattering problems. The numerical results presented in this section were produced on the basis of the FHM methodology described in the previous sections in conjunction with the high-order frequency-domain solvers~\cite{bruno2020chebyshev,garza2021high} (but, for simplicity and definiteness, without use of frequency-domain acceleration methods~\cite{cheng2006wideband, michielssen1994multilevel, bleszynski1996aim, phillips1997precorrected, bruno2001fast, engquist2007fast, bauinger2021interpolated}). As our first example, Figure~\ref{fig:glider} presents an illustration concerning the scattering of an acoustic plane-wave by a simple aircraft. Computational cost comparisons have been performed~\cite{Anderson:20} showing highly competitive results: significant CPU and memory cost savings were observed relative to two prominent solvers proposed recently (one using time-stepping for boundary integral equations and one using the frequency-time convolution-quadrature hybrid method). In short, improvements in computing times and memory by factors of 10s and even 100s were observed, even for short-time simulations. It is expected that even more significant improvement factors would be observed for longer time simulations.

We now consider these two comparisons in detail. The first example concerns a scattering problem involving a sphere of radius $1.6$. The sphere is illuminated by a plane wave \( u^{\textit{inc}}(\mathbf{r}, t) = -a(t - \widehat{\mathbf{k}} \cdot \mathbf{r}) \), where the signal function \( a(t) \) is defined by \( a(t) = 5e^{-(t-6)^2/2} \). The frequency domain was truncated (bandlimited) to the interval \([-W, W]\) with \( W = 6.5 \), and the problem was discretized using 41 equally spaced frequencies \(\omega\) in \([0, W]\). Approximate solutions to the integral equation~\eqref{CFIE_direct} for each frequency were computed using the iterative solver GMRES with un-accelerated forward maps and with a relative residual tolerance of \( 10^{-8} \). The row labeled "This work" in Table~\ref{table} displays maximum solution error, wall time, and memory usage. In particular, approximately four minutes of wall time and 1.2 GB of memory were sufficient to yield a solution with a maximum error around \( 10^{-7} \) relative to a Mie series solution at an observation point $0.2$ units from the scatterer. As discussed in~\cite{Anderson:20}, this experiment is comparable to certain cruller scattering configurations considered in~\cite{Barnett:20}; a comparison between the two sets of results illustrates the favorable performance characteristics of the FHM approach, even for short periods of simulation time.

In a second example we consider a problem of the scattering of a wide-band signal from a unit sphere, given by \( u^{\textit{inc}}(\mathbf{r}, t) = -0.33 \sum_{i=1}^3 \exp\left[\frac{(t - \mathbf{e}_i \cdot \mathbf{r} - 6 \sigma - 1)^2}{\sigma^2}\right] \), where \(\mathbf{e}_1 = (1, 0, 0)\), \(\mathbf{e}_2 = (0, 1, 0)\), \(\mathbf{e}_3 = (0, 0, 1)\), and \(\sigma = 0.1\). The solution achieves an absolute error of \( 2.2 \cdot 10^{-4} \), verified against a multi-incidence Mie series solution at observation point \(\mathbf{r} = (2.5, 0, 0)\). The frequency domain was truncated to \([-W, W]\) with \(W = 45\) and discretized using 91 equally spaced frequencies in \([0, W]\). Solutions to the integral equation~\eqref{CFIE_direct} at each frequency were computed once again using the GMRES iterative solver (without frequency domain acceleration), but now using a relative tolerance of \(10^{-4}\). Performance metrics in Table~\ref{table} illustrate the required computational cost incurred using a 24-core computer system. This test-case was previously considered in
the most recent high-performance implementation~\cite{banjai2014fast} of the convolution quadrature method,
including fast-multipole and $\mathcal{H}$-matrix acceleration. Note that the even in absence of frequency-domain acceleration the proposed method compares favorably with the cited accelerated implementations of the convolution quadrature approach.
\begin{table}[h]
\begin{minipage}[t]{0.49\hsize}\centering
   {\footnotesize
       {\renewcommand{\arraystretch}{1.3}
       \begin{tabular}{|c|c|c|c|}
           \hline
           {---} & {$||e||_\infty$} & {Time} & {Mem.}\\\hline
           This work & $1.6\cdot10^{-7}$  & $4.1$ & $1.2$\\\hline
           Ref.~\cite{Barnett:20} & $\approx10^{-7}$  & $101.75$  & $290$\\\hline
       \end{tabular}}}
\end{minipage}\hfill
\begin{minipage}[t]{0.49\hsize}\centering
{\footnotesize
{\renewcommand{\arraystretch}{1.3}
       \begin{tabular}{|c|c|c|c|}
         \hline
           {---} & {$||e||_\infty$} & {Time} & {Mem.}\\\hline
           This work  & $2.2\cdot10^{-4}$  & $4.3$ & $1.6$\\\hline
           Ref.~\cite{banjai2014fast}  & $2.1\cdot10^{-3}$ & $40.1$ & $56.8$\\\hline
       \end{tabular}}}
\end{minipage}
\caption{Comparison FHM results with results
         in~\cite{Barnett:20} and~\cite{banjai2014fast}. ``This work'' data corresponds to FHM runs
         on a $24$-core computer with Sandy Bridge microarchitecture; the CPUs used in~\cite{banjai2014fast} (resp.~\cite{Barnett:20}) were of a slightly older (resp.\ slightly newer) generation. The ``Mem.'' columns report memory usage in GB. Left: reported times are wall times (in minutes). Right: reported times are CPU core-hours.}\label{table}
\end{table}

\begin{figure}[h!]
\begin{center}
  \includegraphics[width = 1\textwidth]{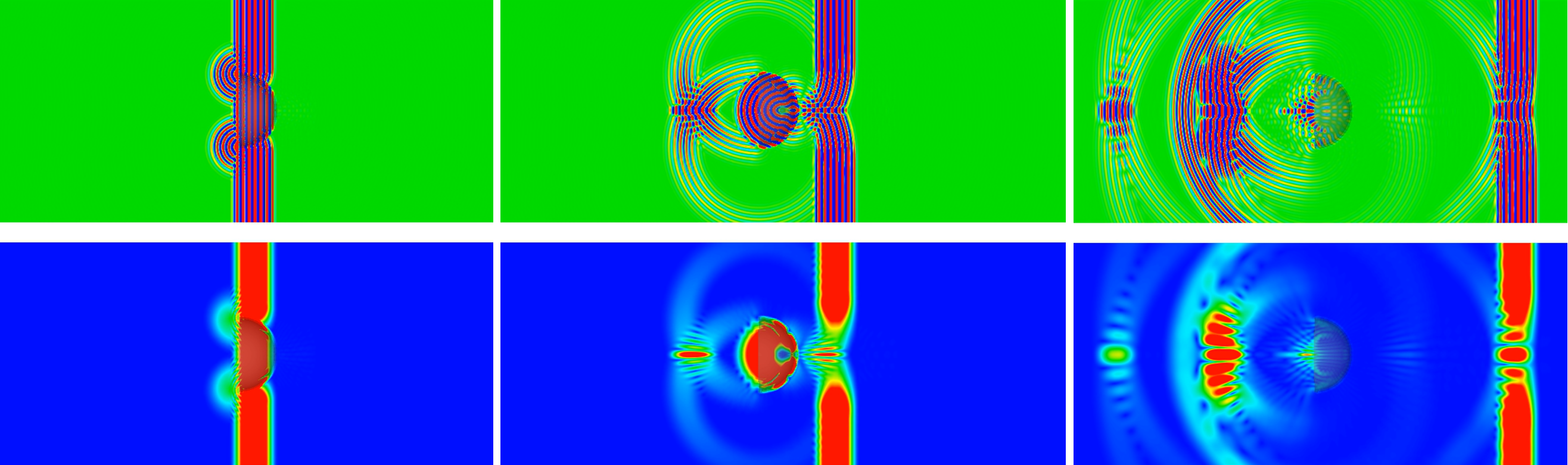}
  \caption{\small Demonstration of the FHM for the Maxwell equations: scattering of a
    time-domain signal by a hemispherical concavity. Upper time
    sequence: real part of the $y$-component of the total electric
    field. Lower time
    sequence: magnitude of the total electric
    field. Credit: Bruno, Tanushev and Voss, unpublished.
    \vspace{-0.3cm}}\label{bruno_tanushev_voss}
\end{center}
\end{figure}

\begin{figure}[h!]
\begin{center}
  \includegraphics[width = 0.7\textwidth]{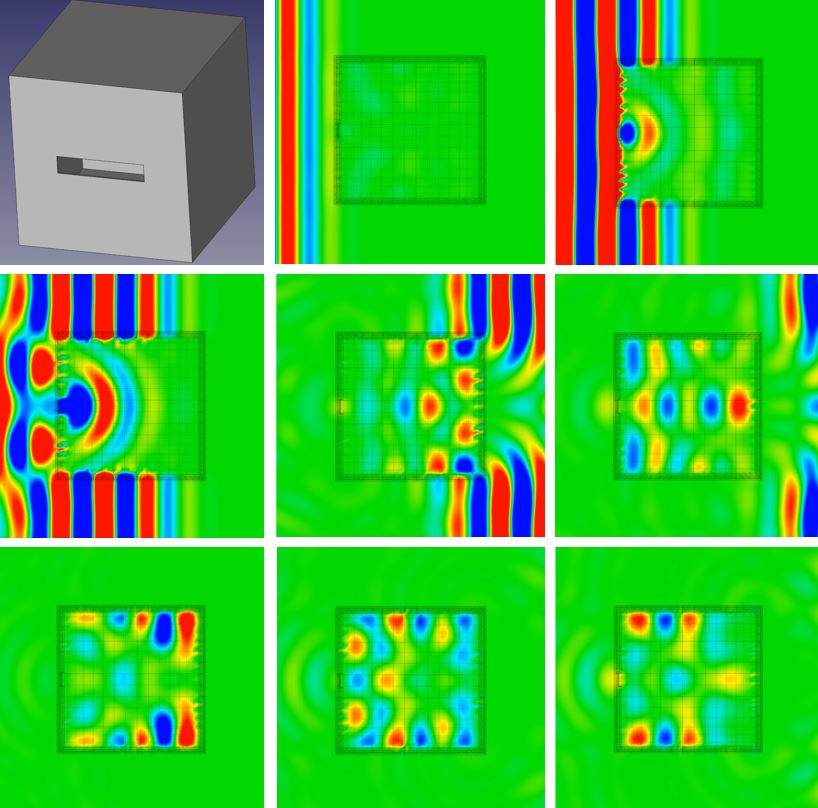}
  \caption{\small Electromagnetic time domain resonances in a slotted cube---an idealized cavity structure of the type arising in field-concentration test cases in areas such as, e.g., electromagnetic compatibility. Upper left: open cavity structure considered. Remaining images: time sequence of events as an incident field impinges on the cavity entrance. As a result of the propagation and scattering events the field is split into two portions, one within the cavity and one outside the cavity, that eventually passes the structure completely. The interior resonant modes continue to operate for long periods of time, allowing for the release of electromagnetic energy on every period of the interior modal field. Credit: Bruno, Tanushev and Voss, unpublished.}\label{bruno_tanushev_voss_2}
\end{center}
\end{figure}

Figures~\ref{bruno_tanushev_voss} and~\ref{bruno_tanushev_voss_2} demonstrate the FHM in the context of 3D time-domain electromagnetic problems. The advantages are similar to those observed in the scalar case. Here we highlight two main characteristics: 1)~the approach can be implemented with any frequency-domain solver at hand (for example, an FHM implementation based on the frequency-domain Method of Moments would also be entirely viable); and, 2)~The method provides dispersion-free high-order accurate propagation to arbitrary simulation times and for arbitrary incident signals of comparable center-frequency and bandwidth on the basis of a fixed set of frequency-domain problems posed on the given domain.


\begin{figure}
    \includegraphics[width=\textwidth]{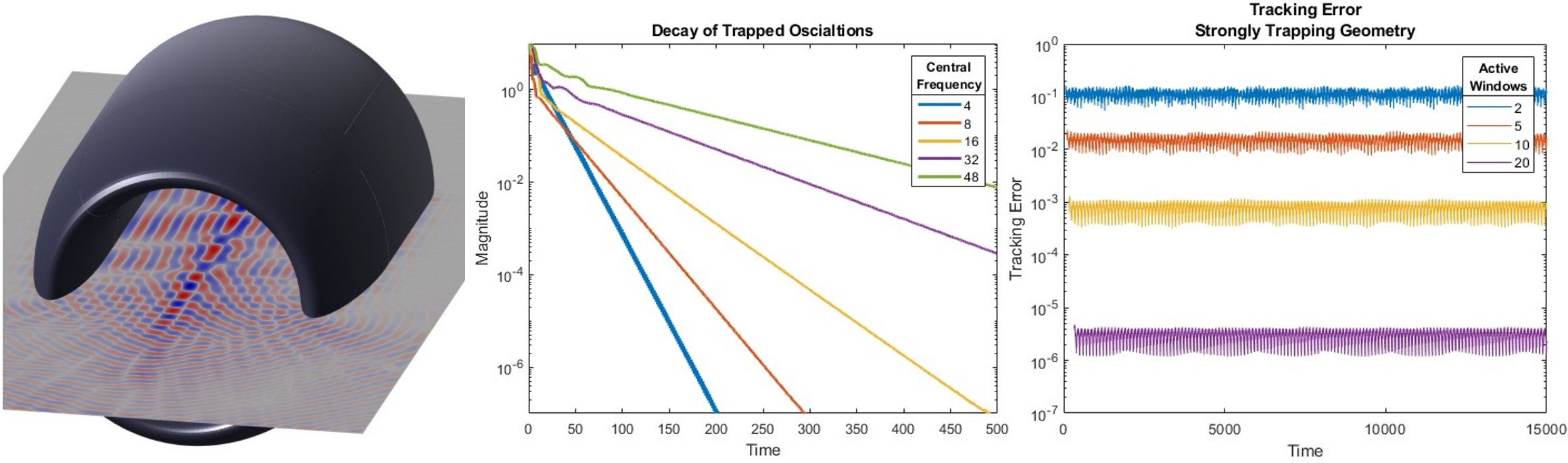}
   \caption{Left and Center: Scattering of a point source from an aircraft nacelle model. The wideband point source is located at a point $\mathbf{r}_0$ within the cavity at distance $\delta$ (equal to $5\%$ of the radius of the opening) above and to the right of the nacelle centerline, and approximately 1/3 of the way along the centerline from the opening of the cavity, with frequency content given by a Gaussian amplitude function distribution. Left: Scattered field along a horizontal plane with center frequency for which the nacelle opening is 16 wavelengths in diameter. Center: Observed decay rates of the maximum magnitude of the scattered field in a cylindrical region of radius equal to double the nacelle opening diameter around the centerline and for center frequencies corresponding to wavelengths equal to $0.7854$, $0.3927$, $0.1963$, $0.0982$ and $0.0654$ nacelle-opening diameters, respectively. Right: ``Tracking Error'' (equal to the absolute-value quantity on the left-hand side of equation~\eqref{eq:eq:ptwise_error}), for a nacelle scattering experiment under an incident chirp field of long duration and with frequencies in the range $1\leq \omega \leq 25$, for various values of the numbers  $M = m_f - m_i$ of active windows used.}
    \label{fig:nacelle}
\end{figure}

It is interesting to examine the character of the FHM algorithm in the contexts of applications to signals of long duration and to problems of long time propagation (see e.g.\ the right panel of Figure~\ref{fig:nacelle}). Indeed, once the necessary (fixed) set of frequency-domain problems are obtained, they can can be used to produce windowed solutions $u_k(\mathbf{r}, t)$, for arbitrarily large sampling values of $t$, at an $\mathcal{O}(1)$ cost. Further, even though the solution $u$ is given by the sum~\eqref{eq:uk_sum}, $u(\mathbf{r}, t) = \sum_{k=1}^K u_k(\mathbf{r}, t)$, which in principle requires an $\mathcal{O}(T)$ cost for large $T$ (since, clearly, $K = \mathcal{O}(T)$), this cost may often be reduced to an optimal $\mathcal{O}(1)$ cost for any fixed but arbitrarily large value of $t$---as it follows from emerging theoretical and algorithmic contributions~\cite{Anderson:24plus,anderson2020bootstrap}. In detail, these references show that, in many cases, for a given bounded region $\mathcal{R}$ around the scatterer, for a given time interval $T_1 \leq t \leq T_2$, and for any given error tolerance $\varepsilon_{\mathrm{tol}}$, a fixed number $M$ of solutions, indexed by $k$ with $m_i\leq k \leq m_f$ ($M = m_f - m_i + 1$), suffices to ensure that the error in the approximation
    \begin{equation}\label{eq:eq:ptwise_error}
        u(\mathbf{r}, t) \approx \sum_{k=m_i}^{m_f} u_k(\mathbf{r}, t),
    \end{equation}
which only contains an $\mathcal{O}(1)$ number of terms,    is less than $\varepsilon_{\mathrm{tol}}$ for any time $t$ in the given time interval. Further, the number $M$ of necessary windowed solutions depends on the desired accuracy $\varepsilon_{\mathrm{tol}}$ and the length of time of interest $T_2 - T_1$, but is otherwise independent of the time $t$ considered---whether large or small. 

As a demonstration of the robustness of the truncation approach just outlined, Figure~\ref{fig:nacelle} presents results of computational experiments for an airplane nacelle-like scattering model which gives rise to a significant amount of wave trapping. The left panel of that figure presents the nacelle geometry  along with the scattered field resulting from a point-source signal emanating from a point within the nacelle cavity. Per the center panel of that figure, the resulting scattered oscillations in the cavity exhibit exponentially fast decay over time after the source is turned off (with slower exponential decay rates for higher frequency signals). The right panel illustrates the truncation methodology used. For this purpose a highly variable chirp incident signal was used, with frequencies of up to approximately $25$, partitioned into one thousand overlapping windows. In view of the error in the right panel, it is apparent that, as mentioned above, only a finite subset of active windows $m_i\leq k\leq m_f$ are necessary at any given time to produce accurate wave scattering solutions with a prescribed accuracy.

\section{Frequency-time hybrid solver for interior domains}

The Fast Hybrid Method just described cannot be applied to problems posed on a bounded domain $\Omega$ on account of the existence of interior Laplace eigenvalues and associated lack of uniqueness to the interior Helmholtz boundary value problem. To overcome this restriction, a novel multiple scattering Fourier transform approach (MS-FHM) relying on decompositions of the closed boundary $\Gamma = \partial \Omega$ into sets of overlapping open surfaces (henceforth referred to as patches) has been developed~\cite{BrunoYin24,PanBaoYinBruno24}; implementations have only thus far been provided in the 2D case where the patches are open arcs. In the first variant of the method, only two such overlapping patches are used while in the second variant arbitrary numbers of patches can be utilized. Rigorous mathematical theorems presented in~\cite{BrunoYin24,PanBaoYinBruno24} show that for any given time the multiple scattering approach yields an exact representation of the solution of the interior problem~\eqref{w_eq} in terms of patch solutions. Importantly, unlike the interior problem, the constituent single patch problems do not suffer from nonuniqueness and, provided the patches are selected adequately (so that they do not individually give rise to trapping behavior), the solution of the corresponding frequency-domain problems require small iteration numbers in an iterative linear algebra solver. Further, the lack of trapping reduces the oscillatory character of the frequency-domain solutions, as a function of frequency, and thus limits the number of required frequency discretization points by the algorithm. In practical terms, as demonstrated in the following sections, the method produces highly accurate interior solutions even for high frequencies and for long time durations. 

\subsection{Variant I: Two-patch MS-FHM}
\label{sec:4.1}
\begin{figure}[htb]
\centering
\begin{tabular}{cc}
\includegraphics[scale=0.4]{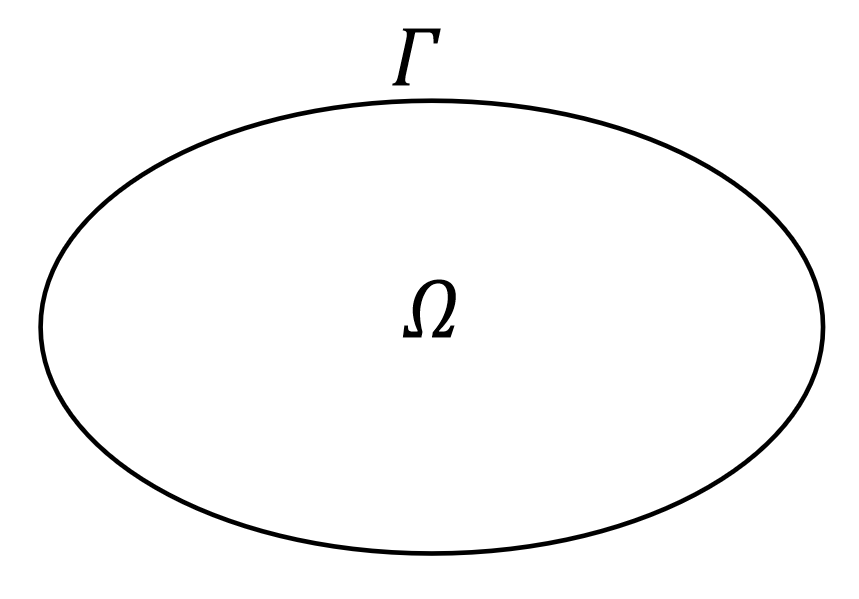} &
\includegraphics[scale=0.4]{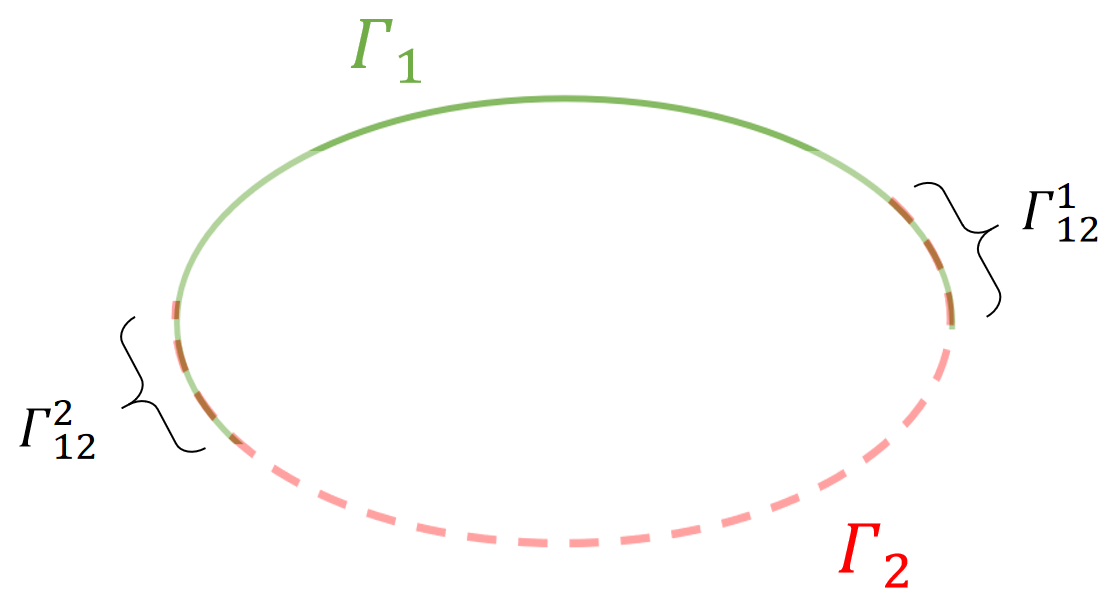} \\
(a) & (b) 
\end{tabular}
\caption{Decomposition of closed boundaries $\Gamma$ depicted in (a) into two
  overlapping open arcs (b).}
\label{MSmodel1}
\end{figure}

The two-patch MS-FHM illustrates some of the key ideas in the context of a simple closed boundary $\Gamma$ which can be decomposed into two overlapping open arcs $\Gamma_1$ and $\Gamma_2$, see Figure~\ref{MSmodel1}; we refer to the connected overlapping regions as $\Gamma_{12}^1$ and $\Gamma_{12}^2$ and set $\Gamma_{12} = \Gamma_{12}^1 \cup \Gamma_{12}^2$. Denoting $\Omega_j = \mathbb{R}^2\backslash \Gamma_j$ ($j=1,2$), we consider the following wave equation problems:
\begin{subequations}\label{w_eq_MS1}
    \begin{align}
        \frac{\partial^2 w_j}{\partial t^2}(\mathbf{r}, t)& - c^2\Delta w_j
            (\mathbf{r}, t) = 0,\quad\mathbf{r} \in \Omega_j,\label{w_eq_MS1_a}\\
        w_j(\mathbf{r},0) &= \frac{\partial w_j}{\partial t}(\mathbf{r}, 0)
            = 0,\quad\mathbf{r} \in \Omega,\label{w_eq_MS1_b}\\
        w_j(\mathbf{r}, t) &=  g_j(\mathbf{r}, t)
            \quad\mbox{for}\quad(\mathbf{r},t)\in\Gamma_j\times [0,+\infty).\label{w_eq_MS1_c}
    \end{align}
\end{subequations}
As described in what follows, the solution $u$ of the interior wave-equation problem up to a given time $t$ can be produced as the sum of a sequence of multiple scattering solutions (\ref{w_eq_MS1}) with adequately chosen right-hand sides $g_j$. 

In detail, introducing the iteration parameter $m\in\mathbb{N}$, we set
\begin{equation}\label{index_shift}
  \quad j = j(m)=  \begin{cases}1&, \quad m = 1, 3, 5, \ldots\\
                                2&, \quad m = 2, 4, 6, \ldots
  \end{cases}
\end{equation}
and we inductively define  boundary-condition
functions $f_m(x,t)$ ($m\geq 1$) and associated wave-equation
solutions $v_m(x,t)$ ($m\geq 1$) as follows: For $m\in\mathbb{N}$ we let  $w_{j(m)}(x,t)$ denote the solution of the open-arc  problem~\eqref{w_eq_MS1} with boundary data
  \begin{equation}\label{bdry_m}
    v_m(\mathbf{r}, t) =  f_m(\mathbf{r}, t)\quad\mbox{for}\quad(\mathbf{r},t)\in\Gamma_{j(m)}\times [0,+\infty),\quad (m \in\mathbb{N}),
  \end{equation}
  where the functions $f_{m}(x,t):\Gamma_{j(m)}\times [0,+\infty)\to\mathbb{C}$ are defined inductively via the relations
    \begin{equation}
  \label{criterion0}
  f_1(\mathbf{r},t)= -u^i(\mathbf{r},t)\quad\mbox{on}\quad\Gamma_1,\quad f_2(\mathbf{r},t)= -u^i(\mathbf{r},t)-v_1(\mathbf{r},t)\quad\mbox{on}\quad\Gamma_2,
\end{equation}
and,   \begin{equation}
  \label{criterion2}
  f_m(\mathbf{r},t)=-v_{m-1}(\mathbf{r},t),\quad\mbox{on}\quad\Gamma_{j(m)}, \quad m \geq 3.
\end{equation}
Then Theorem 2.8 in \cite{BrunoYin24} tells us that, up to time $T(M) = (M - 1)\delta_{12} /c$ (where $\delta_{12}$ equals the distance from $\Gamma_1 \setminus \Gamma_{12}$ to $\Gamma_2 \setminus \Gamma_{12})$), the solution $u(\mathbf{r},t)$ of the problem (\ref{w_eq}) can
be produced by means of the $M$-th order multiple-scattering sum
\begin{equation}\label{mult_scatt1}
  u_M(\mathbf{r},t):=\sum_{m=1}^M
  v_{m}(\mathbf{r},t).
\end{equation}
Evidently, the solution $u_M$ includes contributions from the multiple scattering
iterates $v_m(\mathbf{r},t)$ with $m=1,\dots, M$. As in the previous section, however, only a fixed number of these $M + 1$ multiple-scattering solutions $v_m$ need be included in the sum for highly-accurate solutions for a given fixed time interval.

\begin{figure}[htbp]
    \centering
    \includegraphics[width=0.45\textwidth]{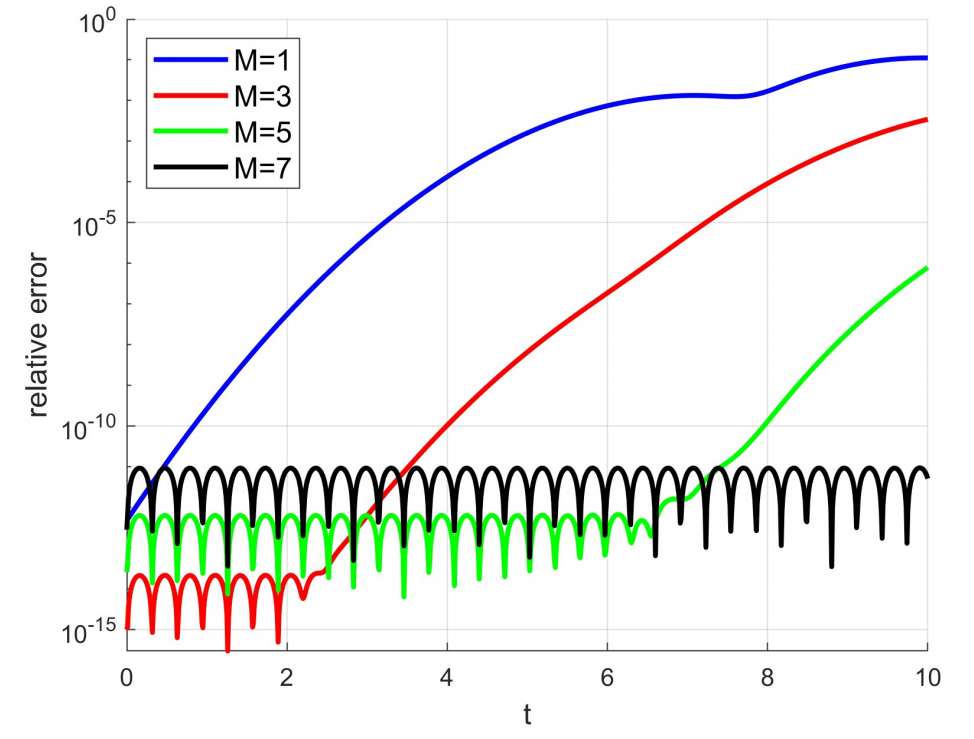}
    \caption{Numerical errors as functions of time $t$ for various values of $M$ for the numerical example of an ellipse consider in Section~\ref{sec:4.1}.}
    \label{fig:twopatchExample}
\end{figure}

As a first example we consider the elliptical boundary $\Gamma=\{\mathbf{r}=(2\cos\theta,\sin\theta)\in\mathbb{R}^2: \theta\in[0,2\pi)\}$ and let the boundary data $b$ be given by $b(\mathbf{r},t)=f(t-t_{lag}-\mathbf{r}\cdot \mathbf{d})$, $\mathbf{r}\in\Gamma$, $t\ge 0$ where 
\begin{equation*}
f(s)=\frac{1}{\sqrt{2\pi}}\exp\{-\frac{s^2}{2\sigma}\}\exp\{-\mathbf{i}\omega_0s\}
\end{equation*}
with $t_{lag}=6$, $\omega_0=15$ and $\mathbf{d}=(1,0)$. The exact solution for this problem is given by $u(\mathbf{r},t)=f(t-t_{lag}-\mathbf{r}\cdot \mathbf{d})$, $\mathbf{r}\in\Omega$, $t\ge 0$. Figure~\ref{fig:twopatchExample} presents the errors (relative to the maximum value of the solution) of the numerical solution at the point $(0.5,0)$ generated by the MS-FHM with sufficiently many frequency discretization points. The results demonstrate the high accuracy of the proposed method and illustrate the equivalence of the multiple-scattering sum (\ref{mult_scatt1}) to the original solution $u(\mathbf{r},t)$ up to time $M \delta_{12}/c$.

\subsection{Variant II: General multi-patch MS-FHM}
\label{sec:4.2}
\begin{figure}[htb]
\centering
\includegraphics[scale=0.5]{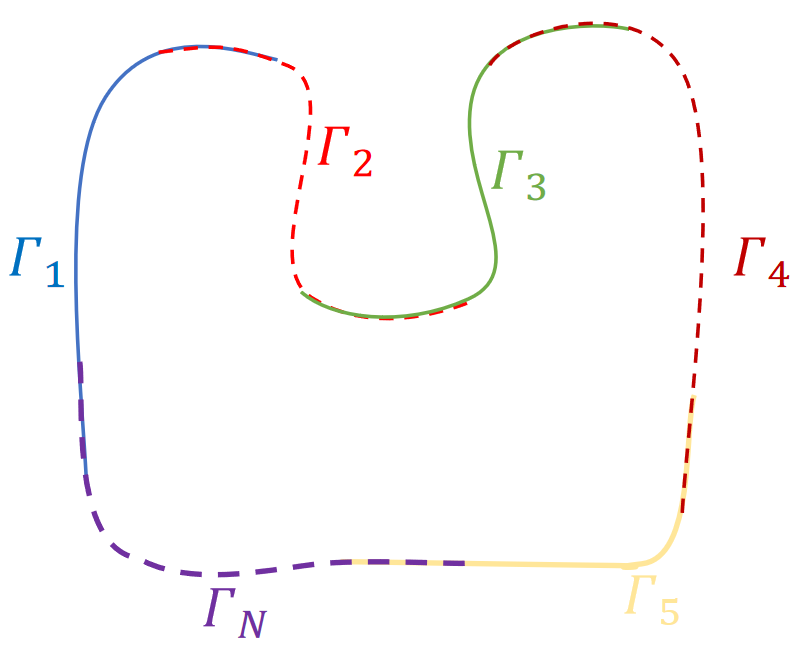}
\caption{An abstract example of the boundary separation strategy using an arbitrary number of overlapping patches  for a closed boundary.}
\label{MSmodel2}
\end{figure}

For a general closed boundary $\Gamma$, a decomposition in an adequate number $N \ge 2$ of patches $\Gamma_j$ may be required to meet the goal that each patch is sufficiently non-trapping. To achieve this goal, the scattering surface
$\Gamma$ is covered by a number $N$ of overlapping patches
$\Gamma_j\subset\partial D$, $j=1,\cdots,N$: $\Gamma=\cup_{j=1}^N \Gamma_j$,  as illustrated in Figure~\ref{MSmodel2}. In what follows we call
$\Gamma_{j,j+1}^{\mathrm{ov}}=\Gamma_j\cap\Gamma_{j+1}$ the overlap of
$\Gamma_j$ and $\Gamma_{j+1}$, we define for convenience
$\Gamma_{N+1}=\Gamma_1$ and $\Gamma_{0}=\Gamma_N$, and we call
$\Gamma_j^{\mathrm{tov}}:=\Gamma_j\backslash
(\Gamma_{j-1,j}^{\mathrm{ov}}\cup \Gamma_{j,j+1}^{\mathrm{ov}})$,
$j=1,\cdots,N$ the portion of $\Gamma_j$ that remains once the regions
of overlap with other patches are ``truncated''. Denoting $\Omega_j = \mathbb{R}^2\backslash \Gamma_j$ ($j=1,\cdots,N$) we then consider the following wave equation problems:
\begin{subequations}\label{w_eq_MS2}
    \begin{align}
        \frac{\partial^2 w_j}{\partial t^2}(\mathbf{r}, t)& - c^2\Delta w_j
            (\mathbf{r}, t) = 0,\quad\mathbf{r} \in \Omega_j,\label{w_eq_MS2_a}\\
        w_j(\mathbf{r},0) &= \frac{\partial w_j}{\partial t}(\mathbf{r}, 0)
            = 0,\quad\mathbf{r} \in \Omega,\label{w_eq_MS2_b}\\
        w_j(\mathbf{r}, t) &=  g_j(\mathbf{r}, t)
            \quad\mbox{for}\quad(\mathbf{r},t)\in\Gamma_j\times [0,+\infty).\label{w_eq_MS2_c}
    \end{align}
\end{subequations}

As in the two-patch FHM, the solution $u$ of the interior wave-equation problem up to a given time $t$ can be produced as the sum of sufficiently many terms in the sequence of multiple scattering solutions generated by the multi-patch MS-FHM embodied in~\eqref{w_eq_MS2} together with the use of windowing functions as described in what follows. 

The appropriate definition of the boundary data $g_j$ requires the introduction of a set of smooth windowing functions that are used to smoothly truncate the time history on each one of the patches used, and thus provide an all-important spatial gap and associated time-span, akin to the time $\delta_{12}/c$ in the two-patch variant of the algorithm, for which the solution produced by each patch may be added without producing incorrect patch-boundary contributions. We thus utilize a smooth partition of unity
$\{\chi_j, j=1,\cdots,N\}$  over $\Gamma$ such that $\chi_j(\mathbf{r})=0$ for $\mathbf{r}\notin\Gamma_j$ and
\begin{equation}
  \label{eq:POU2}
\sum_{j=1}^N\chi_j(\mathbf{r})=1\quad\mbox{for}\quad \mathbf{r}\in\Gamma.
\end{equation}
The multiple scattering solutions are then defined inductively as follows: for $m = 1, 2, 3, \ldots$  define $v_{j,m}(\mathbf{r},t)$ to equal
  the solution $w_j(\mathbf{r},t)$ of the wave equation
  problem~(\ref{w_eq_MS2}) with boundary data
\begin{equation}\label{bdry_int}
v_{j,m}(\mathbf{r},t) = f_{j,m}(\mathbf{r},t)\quad \mbox{for}\quad (\mathbf{r},t)\in \Gamma_j\times [0,+\infty),\quad m \in\mathbb{N},
\end{equation}
for all $j=1,\cdots,N$, where
$f_{j,m}(\mathbf{r},t):\Gamma_{j}\times [0,+\infty)\to\mathbb{C}$ denotes the
functions defined inductively via the relations
\begin{equation}
  \label{criterion_int1}
  f_{j,1}(\mathbf{r},t)= -\chi_j(\mathbf{r})u^i(\mathbf{r},t)\quad (\mathbf{r},t)\in \Gamma_j\times [0,+\infty),
\end{equation}
and, for $m\ge 1$,
\begin{equation}
  \label{criterion_int2}
  f_{j,m+1}(\mathbf{r},t)=-\chi_j(\mathbf{r}) \sum_{k=1, k\ne j}^N\widetilde v_{k,j,m}(\mathbf{r},t), \quad (x,t)\in  \Gamma_j\times [0,+\infty).
\end{equation}
Here, for $j=1,\dots,N$, $(\mathbf{r},t)\in\Gamma_{j}\times [0,+\infty)$,
$k=1,\cdots,N$, $k\ne j$, and $m\ge 1$,
\begin{equation}\label{vtilde}
  \widetilde v_{k,j,m}(\mathbf{r},t)=
\begin{cases}
v_{k,m}(\mathbf{r},t), & \mathbf{r}\in\Gamma_j, k\notin\{j-1, j+1\}  ,\cr
v_{k,m}(\mathbf{r},t), & \mathbf{r}\in\Gamma_j\backslash \Gamma_k, k=j-1\; \mathrm{and}\; k=j+1  , \cr
0, & \mathbf{r}\in\Gamma_{j-1,j}^{\mathrm{ov}}, k=j-1, \cr
0, & \mathbf{r}\in\Gamma_{j,j+1}^{\mathrm{ov}}, k=j+1.
\end{cases}
\end{equation}
A mathematical proof presented in~\cite{PanBaoYinBruno24} shows that the solution $u(\mathbf{r},t)$ of the problem (\ref{w_eq}) can
be produced by means of the $M$-th order multiple-scattering sum
\begin{equation}\label{mult_scatt2}
  u_M(\mathbf{r},t):=\sum_{m=1}^M\sum_{j=1}^N
  v_{j,m}(\mathbf{r},t),
\end{equation}
which includes contributions from the scattering
iterates $v_{j,m}(\mathbf{r},t)$ with $j=1,\cdots,N$,
$m=1,\dots, M$.

\begin{figure}[htbp]
    \centering
    \includegraphics[width=0.45\textwidth]{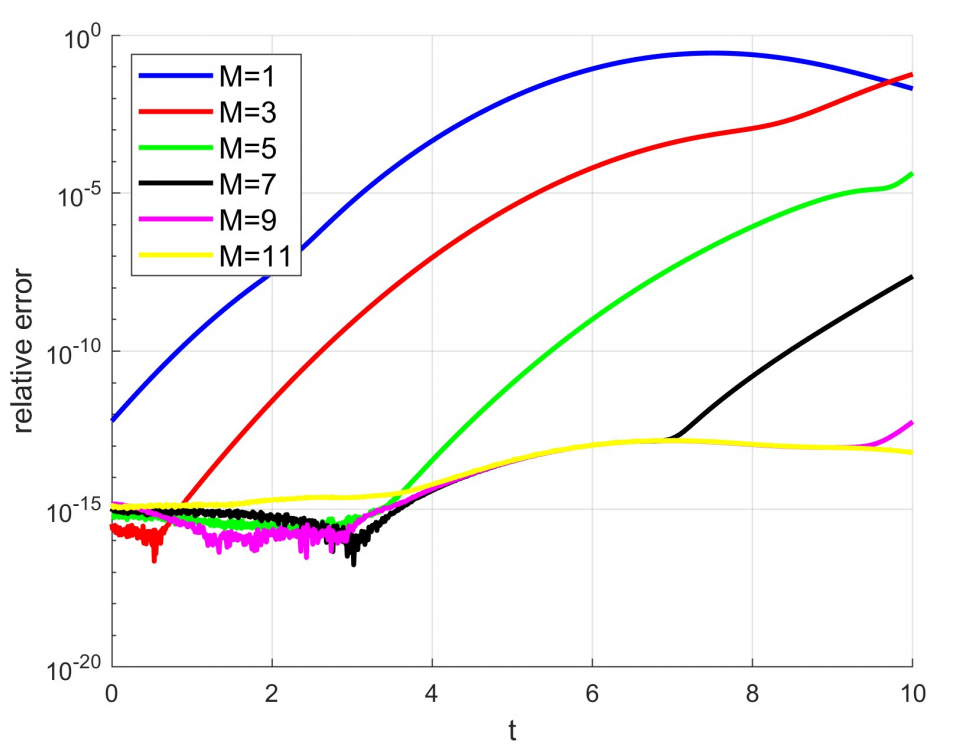}
    \caption{Numerical error as functions of time $t$ for various values of $M$ for the $q=3$-lobed star domain example considered in Section~\ref{sec:4.2} wherein the boundary values are such that the exact solution is known in closed form.}
    \label{fig:multipatchExample1}
\end{figure}

\begin{figure}[htbp]
    \centering
    \includegraphics[width=0.6\textwidth]{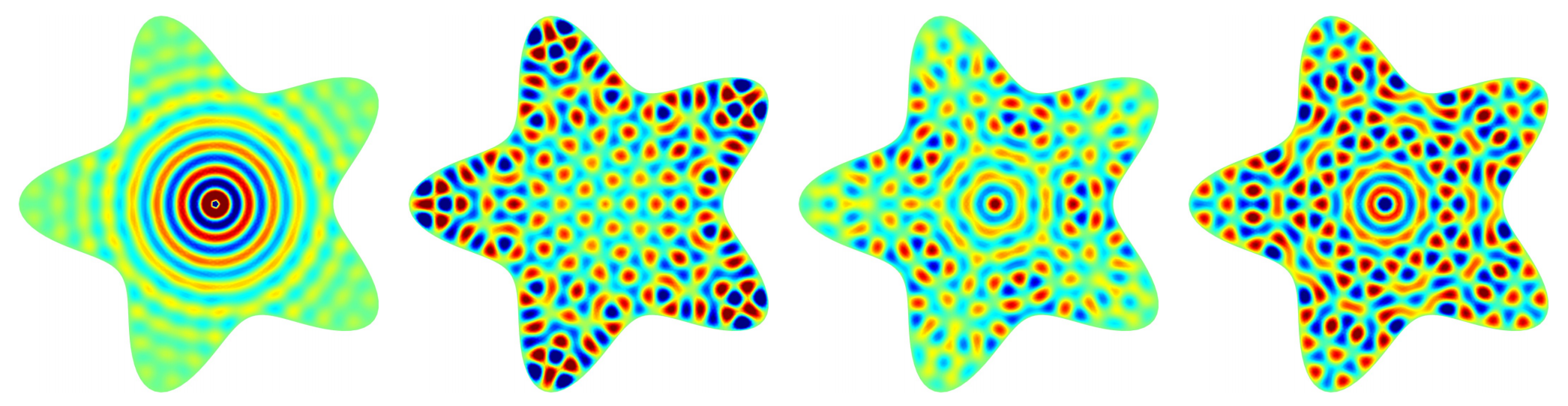}\\
    \includegraphics[width=0.6\textwidth]{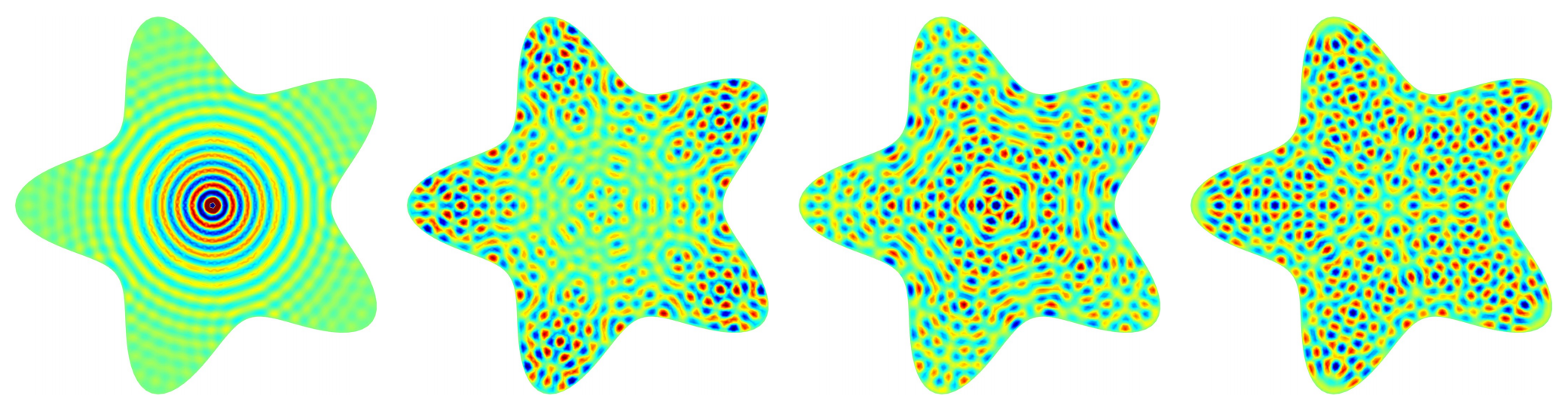}
    \caption{Real part of total fields for the numerical example of $q=5$-lobed star domains considered in Section~\ref{sec:4.2} with point source located at the star center. Upper row: $\omega_0=15$. Lower row: $\omega_0=50$. Fields at times $t=4,6,8,10$ are displayed from left to right in each row.}
    \label{fig:multipatchExample2}
\end{figure}

Letting $\Gamma=\{\mathbf{r}=\rho(\theta)(\cos\theta,\sin\theta)\in\mathbb{R}^2: \rho(\theta)=5(4+\cos(q\theta))^{-1}, \theta\in[0,2\pi)\}$ denote the boundary of the $q$-lobed star domain (depicted in Figure~\ref{fig:multipatchExample2} for $q = 5$) and using the boundary data function introduced in Section~\ref{sec:4.1}, Figure~\ref{fig:multipatchExample1} presents the errors of the numerical solutions at the point $(0.5,0)$, again relative to the maximum value of the solution, for the $q$-lobed star ($q=3$) resulting from the multi-patch MS-FHM with $N=6$ patches. The high accuracy of the proposed method can easily be appreciated, illustrating in particular the equivalence of the multiple-scattering sum (\ref{mult_scatt2}) to the original solution $u(\mathbf{r},t)$ up to a certain time that is proportional to the iteration number $M$.

As a second example we consider a Gaussian-modulated point source $u^i(\mathbf{r},t)$ equal to the inverse Fourier transform of the function
\begin{equation}
\label{pointsource}
U^i(\mathbf{r},\omega)=\frac{5i}{2}H_0^{(1)}(\omega|\mathbf{r}-\mathbf{z}|) e^{-\frac{(\omega-\omega_0)^2}{\sigma^2}}e^{i\omega t_0},
\end{equation}
with $\sigma=2$, $t_0=4$ and $\mathbf{z}=(0,0)$. The resulting boundary data is denoted by $b(\mathbf{r},t)=-u^i(\mathbf{r},t)$, $\mathbf{r}\in\Gamma$, $t\ge0$. Figure~\ref{fig:multipatchExample2} displays the total
fields within a $q=5$-lobed star domain at various times for two choices of $\omega_0$ in~(\ref{pointsource}), each using a total of $M=10$ multiple-scattering iterations and $N=10$ patches.

\section{Applications to dispersive media}\label{sec:dispersive}

This section considers application of Fourier-transform methods to wave propagation in complex media, where, in contrast to the
scattering in homogeneous ``linear'' media  discussed thus far (which are characterized by a linear dispersion relation, e.g.\  $\kappa = \kappa(\omega) = \omega / c$ in the Helmholtz equation~\eqref{helmholtz}, with a constant wavespeed $c$), propagation takes place under a frequency-dependent wavespeed $c=c(\omega)$. A wide variety of methods based on FDTD, FETD, and GFTD have been adapted to situations involving strongly dispersive media, see
e.g.\ \cite{Bleszynski:05}. While some such methods solve a time-domain PDE of
possibly high order, others simulate lossy
media with frequency-dependent loss by means of fractional time-derivative operators, see
e.g.\ references~\cite{Szabo:94, Szabo:95}. These approaches entail substantial memory requirements which, in fact, increase with the final simulation time~\cite{yuan1999simulation}. In contrast, integration of the FHM with problems featuring frequency-dependent media properties is seamless---since a defining feature of the FHM is its ``black-box'' reliance on solution of frequency-domain problems---and it thus incurs fixed memory costs.

This section presents the relevant equations in a simplified, one-dimensional context, with comparison with experimental results; the extension to the multi-dimensional case is direct. For our example we utilize the attenuating media configuration considered in~\cite{Liebler:04}, for which a monochromatic wave $w$ having initial amplitude $w_0$ and frequency
$\omega$ shows exponential attenuation over propagation distance $\Delta x$,
with amplitude following the frequency power-law attenuation relation
\begin{equation}\label{pwr_law}
  \widehat{w}(x + \Delta x) = \widehat{w}_0(x) e^{-\alpha(\omega) \Delta x},\quad \mbox{with }\quad \alpha(\omega) = \frac{\alpha_0}{(2\pi)^\gamma}|\omega|^\gamma,
    \quad 1 \le \gamma \le 2.
\end{equation}
The Fourier transform of the 1D PDE for an idealized homogeneous lossless medium with \emph{constant} wave speed $c_0$,
\begin{equation}\label{w_eq_lossless}
        \frac{\partial w}{\partial t} + c_0 \frac{\partial w}{\partial x} =
        0,\quad w(0, t) = p(t), \quad t > 0,
\end{equation}
is given by the equation
\begin{equation}\label{w_eq_lossless_freq}
    \frac{\partial W(x, \omega)}{\partial x} - i \kappa(\omega) W(x, \omega) = 0,
    \quad\mbox{where}\quad \kappa(\omega) = \beta_0 = \omega/c_0,
\end{equation}
which admits plane wave solutions of the form
$e^{i(\kappa(\omega)x - \omega t)}$.  Following~\cite{Szabo:94}, in order to model attenuation, the dispersion
relation $\kappa(\omega)$ is modified: the resulting dispersion relation  in presence of attenuation must satisfy the 
Kramers-Kronig relations (ensuring analyticity and hence causality of the
waves) which yields~\cite{Szabo:95}
\begin{equation}\label{dispersion_relation}
    \kappa(\omega) = \beta_0 + \beta'(\omega) + \i\alpha(\omega) = \beta_0 +
    \i L^t_\gamma(\omega),
\end{equation}
where $L^t_\gamma(\omega) = -\i \beta'(\omega) + \alpha(\omega) $ represents the change in dispersion relation due to the power-law attenuation in~\eqref{pwr_law}, and the relative dispersion $\beta'(\omega)$ is given by
\begin{equation}\label{relative_dispersion}
    \beta'(\omega) = -\frac{\alpha_0}{(2\pi)^\gamma} \cot((\gamma+1)\pi/2) \omega
    |\omega|^{\gamma-1}.
\end{equation}
Using~\eqref{dispersion_relation} the modified time-domain version of~\eqref{w_eq_lossless_freq} including absorption is given by the causal time-domain wave equation~\cite{Szabo:94}
\begin{subequations}\label{szabo_time_domain_w_eq}
    \begin{align}
        \frac{\partial w(x,t)}{\partial x} + \frac{1}{c_0} \frac{\partial w(x,
        t)}{\partial t} L_\gamma(t) * w(x, t) &= 0,\\
        w(0, t) = p(t), \quad t \ge 0,
    \end{align}
\end{subequations}
Reference~\cite{Liebler:04} presents a modified FDTD-type numerical method for the solution of this equation that includes a strategy for treatment of the computationally challenging convolution term $L_\gamma(t) * w(x, t)$. In that approach the convolution integral is handled on the basis of highly complex recursion strategies that rely on moment-fitting to generate approximations of the time-history in a least squares sense. Validation of numerical simulations against experimental data is presented in~\cite{Liebler:04} for a chirp signal propagating in a lossy medium (castor
oil), between two reference points P1 and P2 that are separated by $1$ cm
(cf.\ \cite[Fig.\ 2]{Liebler:04}).

\begin{figure}
    \center
    \includegraphics[width=0.65\linewidth]{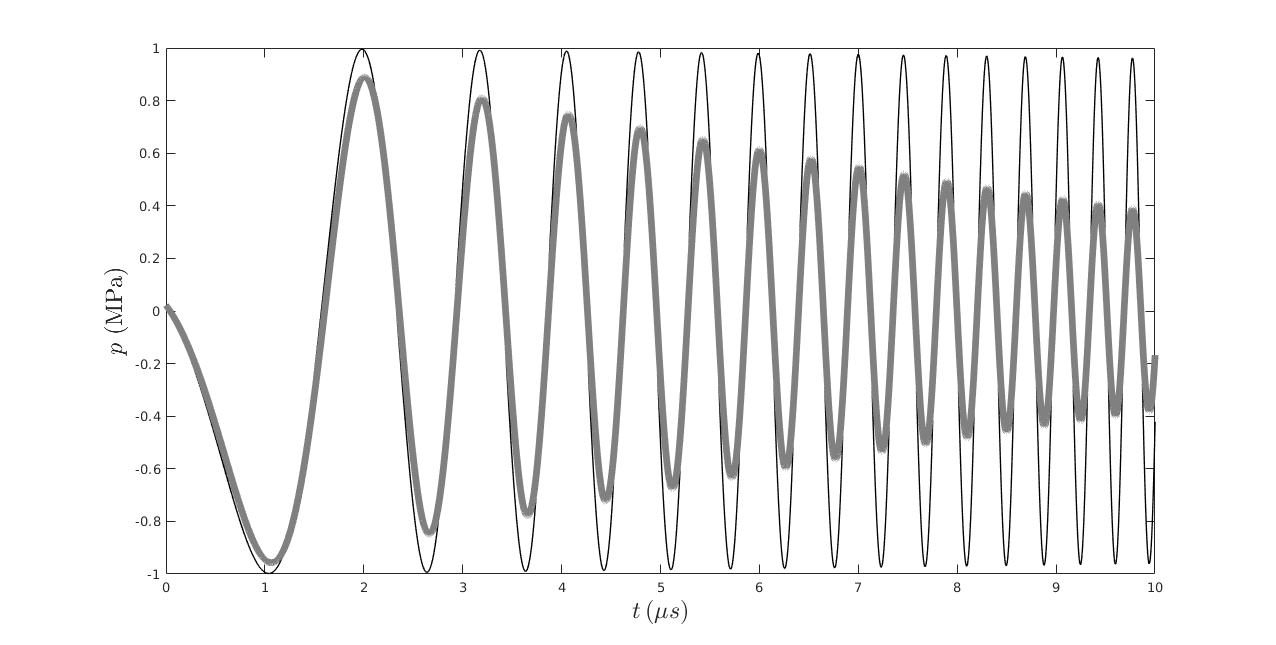}
    \caption{FHM application to a dispersive-medium propagation problem as decribed in the text; compare to~\cite[Fig.\ 3]{Liebler:04}.}
    \label{fig:DispersiveTest}
\end{figure}

We use the FHM to solve the problem~\eqref{szabo_time_domain_w_eq}; that is, we
transform the incoming wave $p(t)$ into discrete Fourier space, on a fixed set
of frequency mesh points $\omega_j\,(1 \le j \le J)$, and then inverse
transform using this fixed set to produce solutions for arbitrarily large time.
Fig.~\ref{fig:DispersiveTest} presents solution values $w(x, t)$ produced by the FHM algorithm
at the previously mentioned reference points P1 and P2.
Following~\cite{Liebler:04}, the two time traces in Figure~\ref{fig:DispersiveTest} are time-shifted by an appropriate time offset $t_o$ so as to illustrate the attenuation by comparison with the incoming disturbance: for a neutral medium a natural choice would be $t_o = \Delta x / c_0$; however, because a generic broadband incident
signal propagates through the medium with a variety of wavespeeds over the active frequency range
$100\,\textrm{kHz}\le f\le 3\,\textrm{MHz}$ due to the variable dispersion relation~\eqref{dispersion_relation}, no single shift is adequate. To
ensure that the highest-frequency components are in phase when comparing the
solutions at P1 and P2, we shift the solution at point P2 by
\begin{equation}\label{offset}
    t_o = \frac{\Delta x}{c(\omega_e)} = \frac{\Delta x}{c_0} \left(1 +
    \beta'(\omega_e)/\beta_0\right), \quad\mbox{where}\quad \omega_e = 2\pi f_e, \quad f_e = 3 \mbox{ MHz}.
\end{equation}
The solution presented in Figure~\ref{fig:DispersiveTest} is visually indistinguishable from the solution presented in~\cite[Fig.\
3]{Liebler:04}, which was produced by means of the significantly more complex algorithm mentioned above and presented in that paper.

\section{Conclusions and outlook}
This paper reviewed a set of recently developed strategies, centered on use of novel forward and inverse Fourier-transformation techniques and allied methodologies, including (i)~A smooth decomposition of a potentially long-duration incident field by means of a smooth partition of unity; (ii)~A recentering approach whereby unnecessarily-fast oscillations as a function of frequency that originate from the mere advancement of time are eliminated; as well as, (iii)~A high-frequency quadrature technique for numerical inverse Fourier transformation; and (iv)~A novel multiple-scattering method for the treatment of trapping structures. The overall approach has been demonstrated for scalar and vector (Maxwell) time-domain problems of scattering by complex structures in 2D and 3D. We believe the demonstrated qualities promise much more, including the applicability to realistic, highly-complex structures---whereby e.g.\ complex geometries with arbitrarily-strongly trapping character can be treated efficiently, with high-accuracy and in short computing times.

\section{Acknowledgements}
OB gratefully acknowledges support from NSF under contract DMS-2109831 and AFOSR under contracts FA9550-21-1-0373 and FA9550-25-1-0015. T. Yin acknowledges with thanks support from the Strategic Priority Research Program of the Chinese Academy of Sciences, Grant No. XDB0640000 and NSFC through Grants 12171465 and 12288201.

\bibliographystyle{IEEEtran}
\bibliography{tdie}

\end{document}